\documentclass[preprint2, 10pt]{aastex}

\slugcomment{Resubmitted to the Astronomical Journal, March 9, 2007}

\shorttitle{A search for HI 21cm absorption 
toward the highest redshift radio loud  objects}
\shortauthors{Carilli et al.}

\begin{document}


\title{A search for HI 21cm absorption toward the highest 
redshift ($z \sim 5.2$) radio loud objects}

\author{C.L. Carilli}
\affil{National Radio Astronomy Observatory, Socorro, NM, USA, 87801}
\email{ccarilli@nrao.edu}

\author{Ran Wang}
\affil{National Radio Astronomy Observatory, Socorro, NM, USA, 87801}

\author{M.B. van Hoven}
\affil{Leiden Observatory, PO Box 9513, 2300 RA, Leiden, The Netherlands}

\author{K. Dwarakanath}
\affil{Raman Research Institute, Bangalore 560080, India}

\author{Jayaram N. Chengalur}
\affil{National Centre for Radio Astrophysics, Post Bag 3, 
Ganeshkhind, Pune 411 007, India}

\author{Stuart Wyithe}
\affil{School of Physics, University of Melbourne, Parkville, 
Victoria, Australia}

\begin{abstract}

We have searched for HI 21cm absorption toward the two brightest radio
AGN at high redshift, J0924--2201 at $z = 5.20$, and J0913+5919 at
$z=5.11$, using the Giant Meter Wave Radio Telescope (GMRT).  These
data set a 3$\sigma$ upper limit to absorption of $< 30\%$ at 40 km
s$^{-1}$ resolution for the 30 mJy source J0913+5919, and $< 3\%$ for
the 0.55 Jy source J0924--2201 at 20 km s$^{-1}$ resolution.  For
J0924--2201, limits to broader lines at the few percent level are set
by residual spectral baseline structure. For J0924--2201 the column
density limit per 20 km s $^{-1}$ channel is: N(HI) $< 2.2\times
10^{18} \rm T_s$ cm$^{-2}$ over a velocity range of $-700$ km s$^{-1}$
to $+1180$ km s$^{-1}$ centered on the galaxy redshift determined
through CO emission, assuming a covering factor of one.  For
J0913+5919 the column density limit per 40 km s$^{-1}$ channel is:
N(HI) $< 2.2\times 10^{19} \rm T_s$ cm$^{-2}$ within $\pm 2400$ km
s$^{-1}$ of the optical redshift. These data rule out any cool, high
column density HI clouds within roughly $\pm 1000$ km s$^{-1}$ of the
galaxies, as are often seen in Compact Steep Spectrum radio AGN, or
clouds that might correspond to residual gas left over from cosmic
reionization.

\end{abstract}

\keywords{cosmology; radio - lines; galaxies - high redshift}

\section{Introduction}

Observations of HI 21cm absorption at cosmologically significant
redshifts provide important probes into a number of physically
interesting problems (Kanekar \& Briggs 2004). First, associated
systems probe the immediate environment of the AGN and the ISM in the
host galaxy.  In particular, searches for HI 21cm absorption toward
Compact Symmetric Radio sources (CSOs) show that in the overwhelming
majority of sources ($\ge 80\%$) absorption is detected at optical
depth levels between 4$\%$ and 40$\%$, with line widths ranging from
about 50 to 500 km s$^{-1}$ (Peck et al. 2000).  CSO's are thought to
be young radio jet sources ($< 10^7$ yrs), possibly confined by
a dense ISM in the host galaxy (Conway 2002). The high HI 21cm
detection rate is thought to be due to the fact that the radio
emission is confined to the inner regions of the galaxy (scales $\le
10$ kpc), and hence HI 21cm absorption searches probe the dense
gaseous environs in the centers of the host galaxies.

Second, studies of intervening HI 21cm absorption systems, in concert
with observations of damped Ly$\alpha$ absorption, have been used to
constrain the evolution of the excitation, or spin temperature of the
neutral gas. Results suggest an increasing fraction of warm neutral
gas with increasing redshift for damped Ly$\alpha$ systems (Kanekar \&
Chengalur 2003; Carilli et al. 1996). This increase might be due to a
change in the nature of the parent galaxy of the absorption line
system from large spirals at low $z$, to smaller dwarf and irregular
galaxies at higher $z$ (Kanekar \& Chengalur 2003), or it could be a
cosmological geometry effect, relating to the very slowly changing
value of the angular diameter distance for redshifts greater than
unity or so (Curran \& Webb 2006).

And third are the recent observational constraints on the epoch of
reionization (EoR), suggesting cosmic reionization started around
$\sim 11$, with the last neutral structures being etched away around
$z\sim 6$ (Fan, Carilli, \& Keating 2006).  It has been pointed out
that the existence of radio loud AGN within the near edge of cosmic
reionization could be used as sensitive probes of intermediate to
small scale structures in the neutral IGM, through HI 21cm absorption
observations (Carilli, Gnedin, \& Owen 2002; Furlanetto \& Loeb 2002),
as opposed to the very large scale that can be studied in emission.

In this paper we present a search for HI 21cm absorption toward two $z
> 5$ radio sources.  While not quite extending into the EoR, these
sources are the highest redshift bright radio AGN known, and the
redshifts are close enough to the end of reionization (within 200
Myr), that we can explore the possibility of enhanced residual cool
neutral gas content in the vicinity of the host galaxy due to
not-fully complete reionization of the densest regions of the universe
(Gnedin 2000; Paschos \& Norman 2005).

\section{The sources}

The source TN J0924--2201 is a radio galaxy with an optical redshift
of $z = 5.19 \pm 0.01$ (van Breugel et al. 1999). The source was
selected from low frequency radio surveys as an 'ultra steep spectrum'
source, with a flux density at 1.4 GHz of 73 mJy and a spectral index
between 0.37 and 1.4 GHz of --1.63. Near IR imaging has revealed a
faint, clumpy source, while radio imaging shows a compact double with
a separation $\sim 1''$ (= 7 kpc), making this source a CSO.  The
Ly$\alpha$ emission from this galaxy is clearly weaker than is normal
for powerful radio galaxies. van Breugel et al. (1999) suggest that
the likely cause for the abnormally weak Ly$\alpha$ line is associated
Ly$\alpha$ absorption, thereby indicating the presence of neutral HI
on kpc-scales in the host galaxy.  The complex near-IR structure, and
the associated Ly$\alpha$ absorption, make this source a prime
candidate for a forming giant elliptical galaxy within a cosmically
over-dense region (Pentericci et al. 2002). Venemans et al. (2004)
estimate a star formation rate in the host galaxy of $\sim 11.4
M_\odot$ yr$^{-1}$, based on the extended (rest-frame) UV continuum
flux as well as on the Ly$\alpha$ line flux. However, this star
formation rate has not been corrected for reddening, which may be
substantial.  Most recently, Klamer et al. (2005) have detected CO
emission from this galaxy, indicating a large reservoir of molecular
gas ($\sim 10^{11}$ M$_\odot$),
and providing a very accurate host galaxy redshift of $z = 5.202 \pm
0.001$.

The source SDSS J0913+5919 is a luminous quasar at $z = 5.11 \pm 0.02$
(Anderson et al. 2002). Radio imaging at 1.4 GHz and 5 GHz reveals a
radio source with a flux density of 18 mJy at 1.4 GHz and a spectral
index of -0.7 (Petric et al. 2003). The radio source is compact, but
possibly marginally resolved at $1''$ resolution, suggesting this
source is a CSO. The optical spectrum of the source shows a relatively
narrow Ly$\alpha$ emission line for a quasar, again likely due to
strong associated absorption, and again indicating the presence of
neutral HI in the host galaxy (Anderson et al. 2001). The optical
luminosity of the QSO implies an Eddington limited mass to the black
hole $\ge 10^9$ M$_\odot$, which would imply a massive host galaxy
($\ge 10^{11}$ M$_\odot$), based on the black hole mass -- bulge
velocity dispersion relation (Gebhardt et al. 2000).

\section{Observations}

Observations were made in the 230 MHz band of the GMRT.  Absolute flux
density calibration was done using 3C48 and 3C286. The source
J0834+555 was observed for phase calibration for J0913+555, and
J0837-198 was used for phase calibration for J0924--2201.  All data
were acquired using the full array of 30 antennas (projected baselines
up to 18 k$\lambda$), giving a synthesized beam $\sim 20"$ FWHM. The
standard line correlator set-up was employed, using 128 spectral
channels.

J0913+5919 has an optical redshift of $5.11 \pm 0.02$, which puts the
HI 21cm line at a frequency of $232.5 \pm 0.8$ MHz. For this source we
observed two polarizations of 4 MHz bandwidth each, centred on the
expected frequency of the redshifted 21cm line. The 128 spectral
channels then yields a spectral resolution of 31.25 kHz, or 40.6 km
s$^{-1}$. 3C286 was used for bandpass calibration.

For J0924--2201 the host galaxy redshift is $z = 5.202 \pm 0.001$,
placing the 21cm line at a frequency of $229.0 \pm 0.1$ MHz.  An
initial short (2 hour) test observation of this source was made in
September, 2003, to investigate the radio frequency interference (RFI)
environment in the 229 MHz band of the GMRT. Although there was strong
interference over much of the 4 MHz band, by careful flagging we were
able to generate a spectrum of the spectral region immediately
defined by the CO redshift, and interestingly, a potential (3$\sigma$
over a few 40 km s$^{-1}$ channels) absorption line was seen.

We reobserved this source in February 2004, with a longer integration
time (three observations of 7 hours each), and a narrower bandwidth
(2MHz) to verify this line.  Unfortunately, the interference
environment for this second observation was much worse than during the
test observations, with only 1 or 2 late-night/early morning hours per
observation being useful.  Moreover, the observations of the bandpass
calibrators 3C48 and 3C286 were completely wiped-out, and the phase
calibrator 0837--198 had to be used for bandpass and gain calibration,
assuming the flux density derived from the test observations. These
latter (admittedly marginalized) data did not verify the line seen in
the test data.

Given the problems with the second data set, we observed J0924--2201
a third time in November 2005, separating the observations over three
nights, but observing only during the two hours per night
corresponding to the lowest RFI levels, and limiting the velocity
range to just the 1MHz band centered on the CO line redshift in order
to avoid strong RFI at the edge of the wider band.

All data processing was performed using the wide-field imaging and
self-calibration capabilities  in AIPS.  

\section{Data analysis}

For the February 2004 observations of J0924--2201, all baselines
shorter than 2 k$\lambda$ ($\sim 50\%$ of all baselines) were
completely corrupted by wideband RFI, and were therefore flagged.
Also, all data were flagged except for roughly two hours in the middle
of each the night.  For the November 2005 observations all baselines
shorter than 1 k$\lambda$ were flagged.  Flagging of residual
channel-by-channel interference removed $\sim 30\%$ of the
channels of the remaining data.

For both sources wide field images were then generated using standard
wide-field imaging routines in AIPS, and the data were self-calibrated
using the resulting models.  For the first (phase only) self
calibration iteration a model based on the NRAO VLA Sky Survey
was used (Condon et al. 1998).

A general issue in low frequency spectral line imaging is side-lobe
confusion from bright sources in the primary beam. In our case this is
especially a problem for J0913+5919, where side lobe confusion is
caused by two relatively bright sources ($\sim 2.4$ Jy and $\sim 1$
Jy) well within the primary beam. In order to overcome this
difficulty, proper modeling and subtraction of continuum sources is
required. We used a combination of bright source model subtraction
from the spectral line data using the AIPS task 'UVSUB', plus spectral
baseline fitting in the uv-plane ('UVLIN') and in the image-plane
('IMLIN').

For the February 2004 observations of J0924--2201, the bandpass
calibration using only the phase calibrator resulted in significant
residual structure across the band. This was removed using a second
order polynomial fit in IMLIN, leaving a $\sim 2\%$ residual bandpass
ripple across the spectrum. For November 2005 we found that bandpass
calibration using 3C286 was adequate to allow for a simple linear
bandpass fit in the image-plane using IMLIN.

In the case of J0913+5919, RFI  on short baselines was
limited to specific channels, the data could therefore be edited quite
efficiently using the AIPS tasks 'SPFLG', 'UVFLG' and 'FLGIT',
ultimately resulting in the removal of about 35$\%$ of all data.
A linear baseline was adequate for residual continuum subtraction 
with IMLIN.

\section{Results}

Figure 1 shows the continuum image of each source. Both sources are
unresolved at the resolution of the GMRT ($\sim 20''$). J0913+5919 was
found to have a flux density of $30 \pm 3$ mJy at 230MHz, while
J0924--2201 had a flux density at of $0.55\pm0.05$ Jy. In both cases
the errors represent observation-to-observation differences, likely
due to residual errors in the flux density boot-strap process.

The continuum subtracted spectrum of J0913+5919 is shown in Figure 2a.
The rms noise is 3 mJy per 40.7 km s$^{-1}$ channel.  At this
resolution, we find no absorption to a 3$\sigma$ optical depth limit
of 0.3, over a velocity range of $\pm 2400$ km s$^{-1}$ centered on the
optical redshift of $z = 5.11 \pm 0.02$.

There is a broad depression in the J0913+5919 spectrum, covering much
of the velocity range dictated by the optical redshift plus
uncertainty. We have smoothed the spectrum to 164 km s$^{-1}$, and the
result is shown in Figure 2b. The rms per channel is 2 mJy, and the
depression can be seen centered at about 232.3 MHz, or $z = 5.114$,
with a width of about 9 channels, or 1500 km s$^{-1}$. The mean depth
of the depression is about 2 mJy, or just 1 sigma per channel,
although this persists over 9 channels. Note that we have used all of
the channels shown for linear spectral baseline subtraction. If we
exclude the 9 'line' channels from the spectral baseline fit, the mean
depression depth increases to almost 3mJy. We do not consider
this a line detection, but a potential result that requires further
observation.

Figure 3 shows the results for J0924--2201. Figure 3a shows the
wide-band spectrum from February 2004, for which a second order
spectral baseline has been removed using all of the channels shown.
We do not detect any absorption over the full band to an rms level of
11 mJy per 20 km s$^{-1}$ channel, corresponding to a 3$\sigma$
optical depth limit of 6\%. The frequency range covered corresponds to
$-700$ km s$^{-1}$ to $+1180$ km s$^{-1}$, centered on the CO redshift
for the host galaxy. Figure 3b shows the spectrum from November 2005
at 20 km s$^{-1}$ resolution.  Again, no absorption is detected to an
rms level of 6 mJy per 20 km s$^{-1}$ channel, corresponding to a
3$\sigma$ optical depth limit of 3.3\%. The frequency range covered
corresponds to $\pm 460$ km s$^{-1}$. We have combined the two
spectra, weighting by the rms, and the result is show in Figure
3c. The rms per 20 km s$^{-1}$ channel is now 5 mJy, implying a
3$\sigma$ optical depth limit of 2.7\%. Residual spectral baseline
structure limits any broad ($\sim 1000$ km s$^{-1}$) lines to $< 2\%$.

\section{Discussion}

These GMRT observations set $3\sigma$ upper limits to the HI optical
depth toward J0924--2201 of 3\% at 20 km s$^{-1}$ resolution within
$\pm$460 km s$^{-1}$ of the host galaxy redshift, and 6\% over a wider
velocity range of $-700$ km s$^{-1}$ to $+1180$ km s$^{-1}$. The
column density limit per channel is then: N(HI) $< 1.1\times 10^{18}
\rm T_s$ cm$^{-2}$ over the narrow range, and $< 2.2\times 10^{18} \rm
T_s$ cm$^{-2}$ over the broader range, where $\rm T_s$ is the HI spin
temperature. Residual broad spectral baseline structure limits any
broad absorption line to $< 2\%$. The column density limit is then $<
1.8\times 10^{19} \rm T_s$ cm$^{-2}$ for a 500 km s$^{-1}$ line.

For the weaker radio source, J0913+5919, the optical depth limit is
0.3 at 40 km s$^{-1}$ resolution, within $\pm 2400$ km s$^{-1}$ of the
optical redshift. The column density limit per channel is then: N(HI)
$< 2.2\times 10^{19} \rm T_s$ cm$^{-2}$.  There is a broad, weak
depression in the center of the spectrum of about 1500 km s$^{-1}$
width with a depth of about 7\%, although this signal is really only
1$\sigma$ in 9 channels, or roughly 3$\sigma$ averaged over all of the
'line' channels. If real, the implied column density is: N(HI) $\sim
2\times 10^{20} \rm T_s$ cm$^{-2}$.  We consider this an upper limit
to a broad line in this source, although further observations of this
source would be very interesting.

We note that both sources in our study show evidence for an extended
radio continuum source of $\sim 1"$ in size ($\sim 6$ kpc; see section
2). Low HI covering factors of the continuum sources could also lead
to low apparent optical depths.

For J0924--2201, assuming that the very weak Ly$\alpha$ emission line
from the host galaxy is due to high column density (damped) associated
Ly$\alpha$ absorption line system (N(HI)$ \sim 10^{21} $cm$^{-2}$),
and using the HI 21cm limits above, suggests that the HI spin
temperature is relatively high ($>$few hundred K).  For comparison,
Galactic clouds with N(HI) $\ge 10^{21} $cm$^{-2}$ typically have spin
temperature values $\sim 100$ K for the absorbing gas (Dwarakanath,
Carilli \& Goss 2001). For damped Ly$\alpha$ systems at $z = 3$
typical values for $T_{spin}$ on the order of $10^{3}$K are found
(Carilli et al., 1996; Kanekar \& Chengalur 2003). 

These are (by far) the two brightest radio sources known at $z >
5$. The redshifts of these two sources place them within 200 Myr of
the end of reionization ($z \sim 6$), as determined from eg. the
Gunn-Peterson effect toward high z QSOs (Fan et al. 2006). Further,
these luminous AGN are likely associated with denser regions of the
universe, ie. proto-clusters, at the earliest epochs (Venemans et
al. 2004). Given that reionization is likely a process extended in
space and time (Fan, Carilli, Keating 2006), two possibilities exist
for the large scale environments of the systems. There may be a
significant residual population of dense HI clouds associated with the
denser regions of the universe (Gnedin 2000; Paschos \& Norman
2005). In this case, one might expect to see 21 cm absorption not just
by gas in the inner kpc's of the radio source host galaxy, but perhaps
by the densest residual HI clouds surrounding the host galaxy on
scales of tens to hundreds of kpc, left over from the EoR and still
being ionized by the increasing uv background radiation
field. Conversely, these densest regions may ionize earliest, due to
the biased formation of luminous structure, and the associated higher
ionizing radiation field (Wyithe \& Loeb 2006).

Our observations have not detected any absorption by neutral systems
in the IGM surrounding the $z \sim 5.2$ quasars, implying that there
are no residual neutral hydrogen clouds along their lines of
sight. Indeed, we can certainly rule-out any large (kpc-scale), high
column density, cool HI clouds within 1000 to 2000 km s$^{-1}$ along
the lines of site to either of these these sources. This lack of
absorption is qualitatively consistent with expectations based both on
optical observations of high redshift ($z\sim 6$) quasars, and on
theoretical expectations from models of the reionization process. As
seen in the SDSS quasars at $z \sim 6$ (White et al. 2003), there
should be a proximity zone extending to $\sim$ 3000 km s$^{-1}$
blueward of the quasar redshift, within which the IGM is sufficiently
ionized that Ly-alpha photons are transmitted.

On the theoretical side, we would expect the ionizing background
(following reionization) to be much larger than average within 1 MHz
of the quasar redshift. For example, the enhancement of the
ionizing backgound as a function of radius from a massive halo has
been computed by Dijkstra, Lidz, \& Wyithe (2006). When applied to a
quasar, they find that the typical enhancement predicted by this model
at a distance of 1Mpc ranges between factors of 10 and 100, depending
on the quasar host mass. Much of this enhancement is due to clustering
of sources around the quasar, which preferentially form in the
overdense infall region (Wyithe \& Loeb 2005). This large enhancement
in the ionizing background radiation would suppress any absorption
signature close to the quasar.

Recent models describe the process of reionization as patchy, meaning
in part that overlap is completed at different times in different
regions of the IGM. However the time of reionization within a region
of IGM is not expected to be random. Rather, galaxy bias will ensure
that overdense regions are reionized earlier than the average IGM
(e.g. Wyithe \& Loeb 2006). Rare objects, such as the massive halos
that house quasars at $z\sim5.2$, formed preferentially in overdense
regions. It follows that the distribution of overdensities on 1Mpc
scales surrounding these quasars is significantly skewed towards
positive values (Wyithe \& Loeb 2006). We therefore expect
reionization to occur earlier around such objects. As an example,
using the model described in Wyithe \& Loeb (2006), we can calculate
the time at which the IGM within 1Mpc of a quasar is reionized
relative to the average IGM. Even in a case where the quasar flux is
ignored, such a region would only be reionized at a time later than
the average IGM in 5
and current models of reionization, suggest that the immediate infall
regions around bright quasars ($\sim$ a few Mpc) are not the best
sites to search for residual cool neutral gas clouds near the end of
cosmic reionization. Future searches for such primordial HI clouds
toward $z > 6$ radio sources are best done at redshifts a few thousand
km s$^{-1}$ below the AGN host galaxy redshift. 

Currently, the GMRT is the largest area telescope capable of observing
the HI 21cm line into cosmic reionization, ie. at frequencies close
to, or below, about 200 MHz. The interference environment clearly
makes such studies difficult, and any further progress in very high
redshift HI 21cm absorption searches using existing instrumentation will
require discovery of bright ($> 100$mJy) radio sources at redshifts
that happen to put the HI line in the ever decreasing RFI-free parts
of the low frequency spectrum.  However, within a few years telescopes
such as the Mileura Widefield Array, the Low Frequency Array, the
Precision Array for Probing the Epoch of Reionization, and the 21
Centimeter Array, some of which will operate at sites chosen to
minimize terrestrial interference, should allow for the study of
fainter sources (few mJy), over wider frequency ranges (Carilli 2006).

\acknowledgments We acknowledge support from the Max-Planck Society
and the Alexander von Humboldt Foundation through the Max-Planck
Forshungspreise 2005. We thank the referee for useful comments.  The
National Radio Astronomy Observatory is a facility of the National
Science Foundation, operated by Associated Universities, Inc.

\clearpage
\newpage

\begin{figure}[ht]
\includegraphics[width=2.7in]{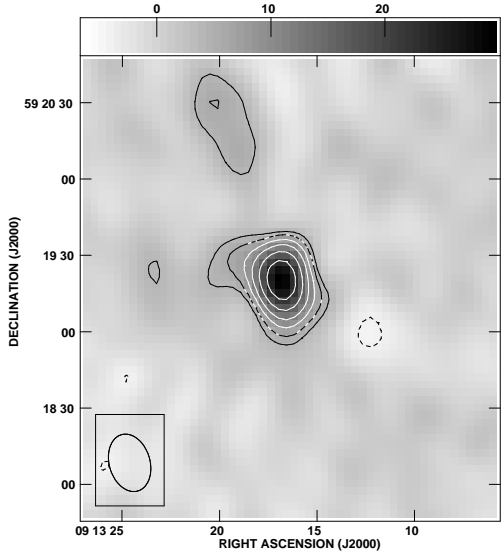}\\
\includegraphics[width=2.7in]{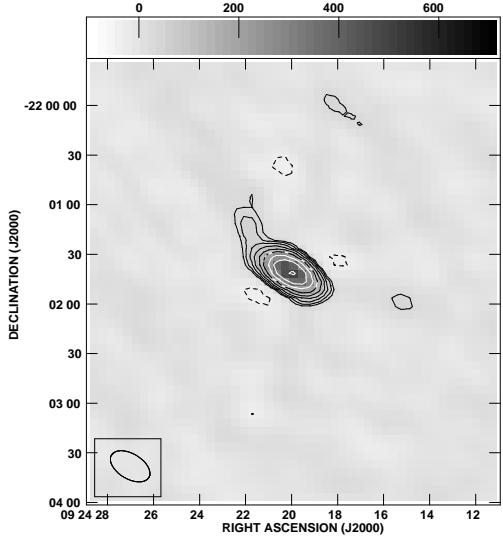}
\caption{ {\bf Upper}: The GMRT continuum image at 232 MHz
of J0913+5919 at $z = 5.11 \pm 0.02$. The source continuum flux density
at this frequency is $30\pm3$mJy. The 
spatial resolution is $23"\times16''$ with major axis
position angle= 16$^o$. The contour levels are a geometric
progression in the square root of two starting at 4 mJy. 
Negative contours are dashed.{\bf Lower}: The GMRT continuum image 
at 229 MHz of J0924-2201 at $z = 5.202 \pm 0.001$. The source continuum 
flux density at this frequency is $0.55\pm0.05$Jy.  The 
spatial resolution is $26"\times15''$ with major axis
position angle=  59$^o$. The contour levels are a geometric
progression in the square root of two starting at 30 mJy. 
Negative contours are dashed.
}
\label{}
\end{figure}

\begin{figure}[ht]
\includegraphics[angle=-90,width=4in]{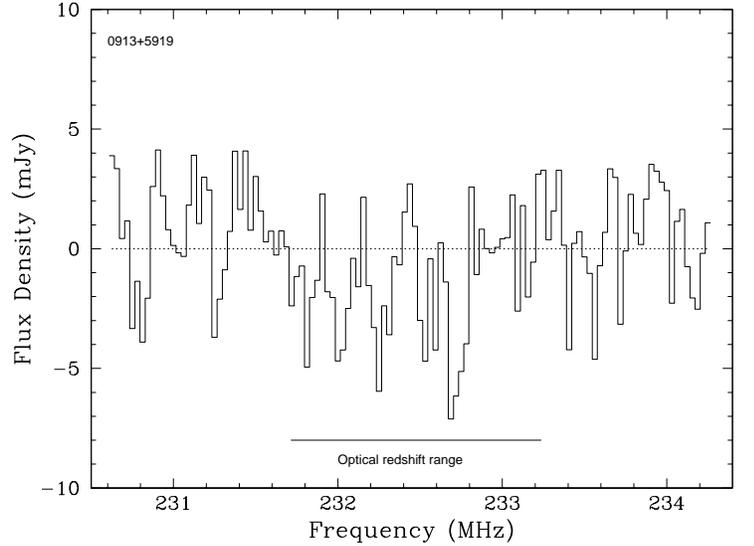}\\
\includegraphics[angle=-90,width=4in]{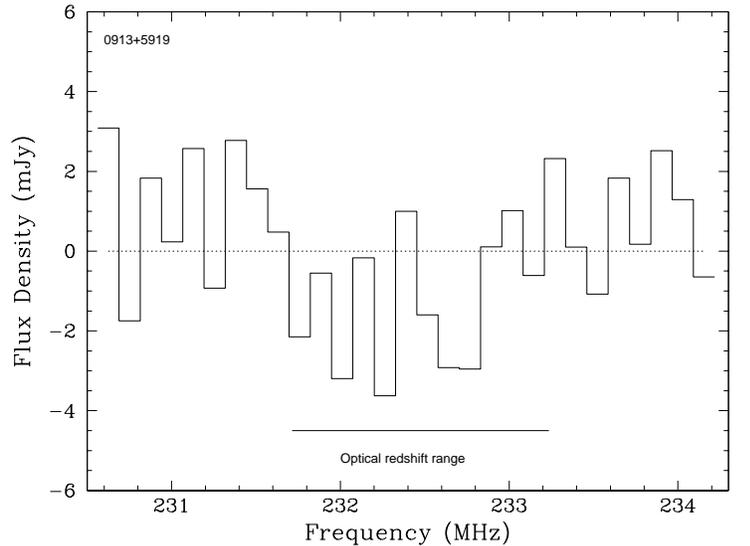}
\caption{ {\bf Upper}: The continuum subtracted 
GMRT spectrum of J0913+5919 at $z = 5.11 \pm 0.02$
at 40 km s$^{-1}$ resolution. The source continuum flux density
at this frequency is $30\pm3$mJy. The horizontal bar shows the
redshift range dictated by the uncertainty in the optical
redshift ($=2000$ km s$^{-1}$). 
{\bf Lower}: The same spectrum, smoothed to 164 km s$^{-1}$ resolution.
}
\label{}
\end{figure}

\clearpage
\newpage

\begin{figure}[ht]
\includegraphics[angle=-90,width=3.2in]{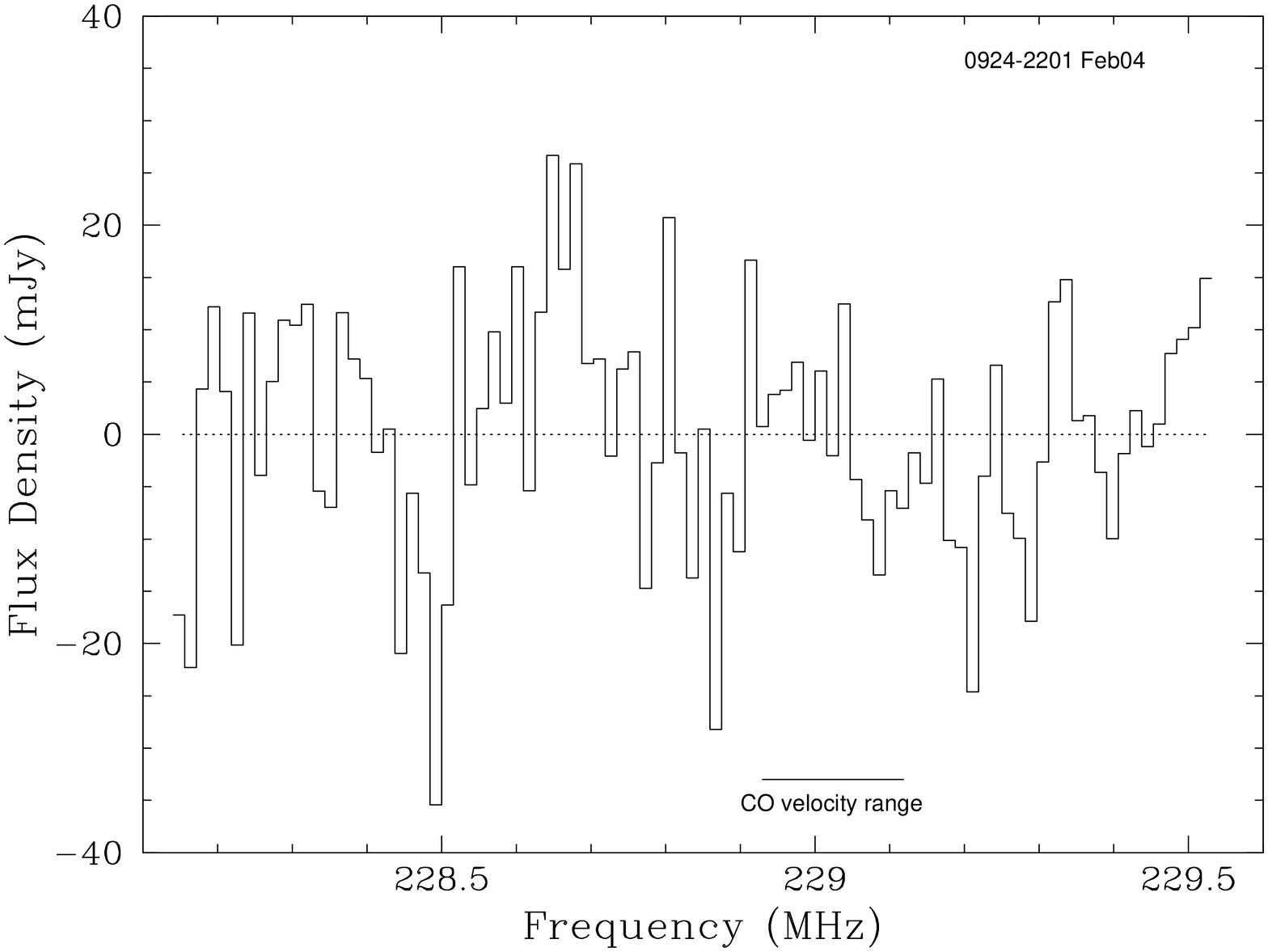}\\
\includegraphics[angle=-90,width=3.2in]{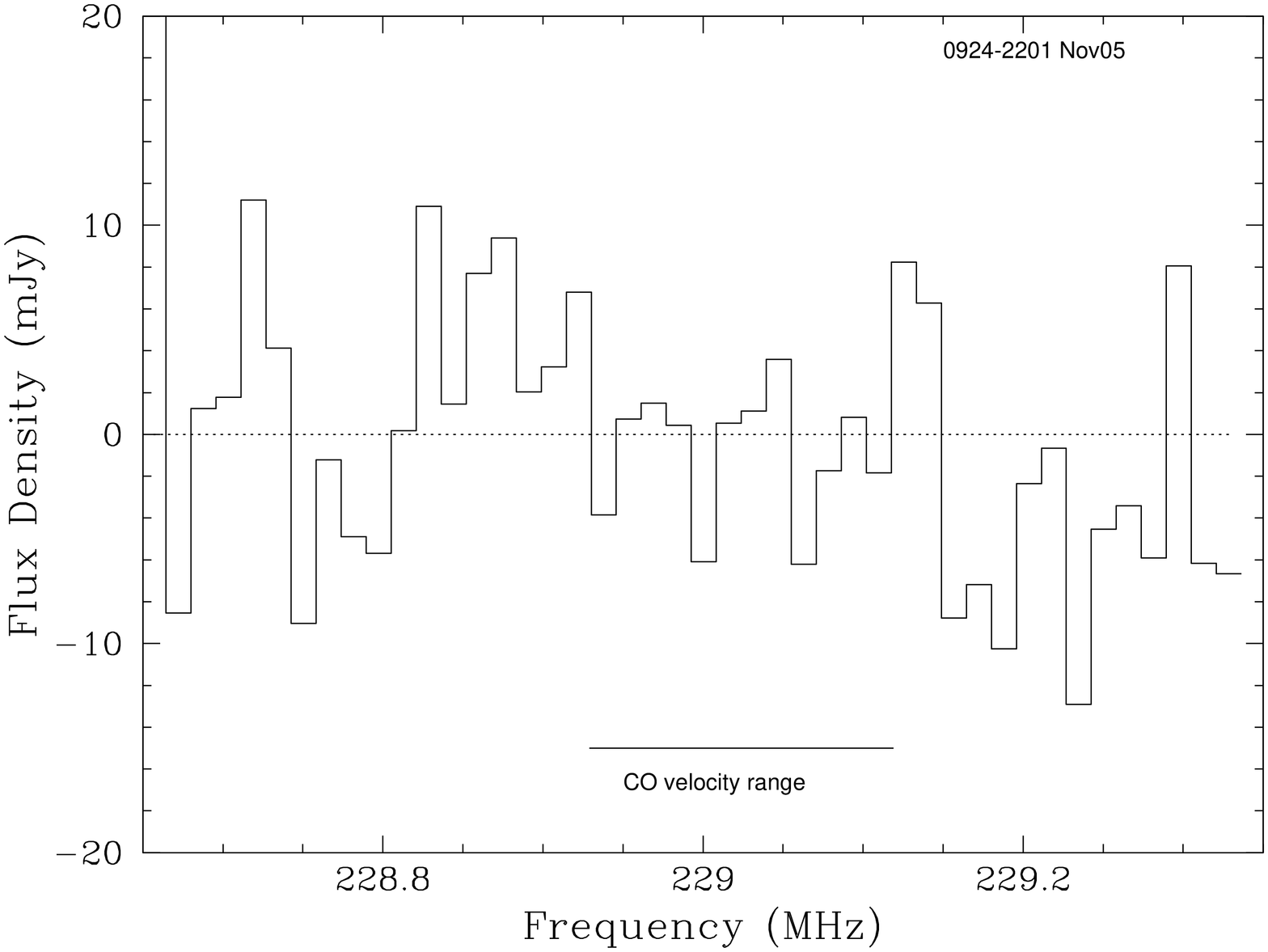}\\
\includegraphics[angle=-90,width=3.2in]{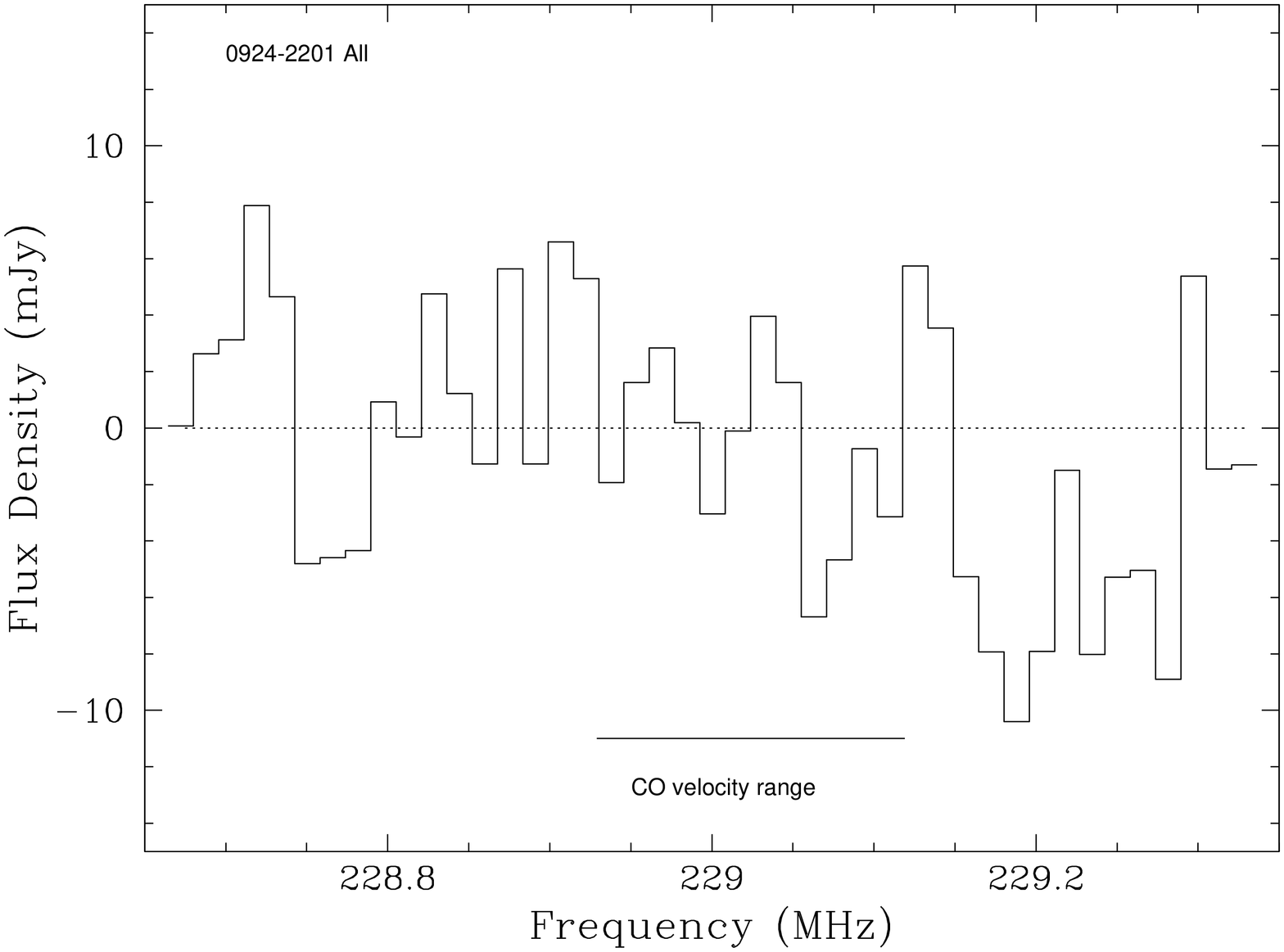}
\caption{ {\bf Upper}: The continuum subtracted 
GMRT spectrum of J0924--2201
at 20 km s$^{-1}$ resolution for data taken in February 2004.
The source continuum flux density at this frequency is $0.55\pm0.05$Jy. 
The host galaxy redshift is $z = 5.202 \pm 0.001$, determined
from CO emission (Klamer et al. 2005). The  horizontal bar shows the
redshift range dictated by the width of the CO emission line
(=250 km s$^{-1}$).
{\bf Middle}: The same as above, but for data taken in November 2005 over
a narrower bandwidth. 
{\bf Lower}: The combined spectrum from November 2005 and February 2004. 
}
\label{}
\end{figure}

\end{document}